\documentclass[]{aa} 
\usepackage{graphicx}
\usepackage{txfonts}
\begin{document}
   \title{Dynamical evolution of titanium, strontium, and yttrium spots on the surface of the HgMn star \object{HD\,11753}\thanks{Based on observations obtained with the CORALIE Echelle Spectrograph on the 1.2-m Euler Swiss telescope, situated at La Silla, Chile}}


   \author{
M.\,Briquet\inst{1}\thanks{Postdoctoral Fellow of the Fund for Scientific Research, Flanders}
\and 
H. Korhonen\inst{2}
\and 
J.F.\,Gonz\'alez\inst{3}
\and 
S.\,Hubrig\inst{4}
\and 
T. Hackman\inst{5}
   }

   \institute{Instituut voor Sterrenkunde, Katholieke Universiteit Leuven, Celestijnenlaan 200 D, B-3001 Leuven, Belgium\\
     \email{maryline@ster.kuleuven.be}
     \and 
     European Southern Observatory, Karl-Schwarzschild-Str 2, D-85748 Garching bei M\"unchen, Germany         
     \and
     Instituto de Ciencias Astronomicas, de la Tierra, y del Espacio (ICATE), 5400 San Juan, Argentina
     \and    
     Astrophysikalisches Institut Potsdam, An der Sternwarte 16, 14482 Potsdam, Germany
     \and   
     Observatory, PO Box 14, FI-00014 University of Helsinki, Finland}

   \date{Received; accepted}

 
  \abstract
   {}
   {We gathered about 100 high-resolution spectra of three typical HgMn (mercury-manganese) stars, \object{HD\,11753}, \object{HD\,53244}, and \object{HD\,221507}, to search for slowly pulsating B-like pulsations and surface inhomogeneous distribution of various chemical elements. }
   {Classical frequency analysis methods were used to detect line profile variability and to determine the variation period. Doppler imaging reconstruction was performed to obtain abundance maps of chemical elements on the stellar surface. }
   {For \object{HD\,11753}, which is the star with the most pronounced variability, distinct spectral line profile changes were detected for Ti, Sr, Y, Zr, and Hg, whereas for \object{HD\,53244} and \object{HD\,221507} the most variable line profiles belong to the elements Hg and Y, respectively. We derived rotation periods for all three stars from the variations of radial velocities and equivalent widths of spectral lines belonging to inhomogeneously distributed elements: P$_{rot}$ (\object{HD\,11753})=9.54\,d, P$_{rot}$ (\object{HD\,53244})=6.16\,d, and P$_{rot}$ (\object{HD\,221507})=1.93\,d. For \object{HD\,11753} the Doppler imaging technique was applied to derive the distribution of the most variable elements Ti, Sr, and Y using two datasets separated by $\sim$65 days. Results of Doppler imaging reconstruction revealed noticeable changes in the surface distributions of \ion{Ti}{ii}, \ion{Sr}{ii}, and \ion{Y}{ii} between the datasets, indicating the hitherto not well understood physical processes in stars with radiative envelopes that cause a rather fast dynamical chemical spot evolution.}
   {}


   \maketitle
%

\section{Introduction}
The mercury-manganese (HgMn) stars constitute a well-defined sub-class of chemically peculiar (CP) stars of the B7--B9 spectral types with $T_{\rm eff}$ between 10\,000 and 15\,000\,K. These stars exhibit marked abundance anomalies of several elements: e.g., overabundances of Hg, Mn, Ga, Y, Cu, Be, P, Bi, Sr, Zr, and deficiencies of He, Al, Zn, Ni, Co. More than two thirds of them belong to spectroscopic binaries (Hubrig \& Mathys \cite{hubrig_mathys95}). They are slow rotators ($\langle v\,\sin i\rangle=29$~km$\,$s$^{-1}$, Abt \cite{abt72}). There is no evidence that they would have strong large-scaled organized magnetic fields. Their elemental overabundances/underabundances are believed to be due to radiatively-driven diffusion and gravitational settling.

In the H-R diagram many HgMn stars are located in the instability strip of the so-called slowly pulsating B (SPB) stars (see De Cat \cite{decat03} for a review on the latter), and the sophisticated models predict that pulsations should also be driven in HgMn stars (Turcotte \& Richard \cite{turcotte_richard05}). Searches for variability in this group of stars have been made mostly photometrically in the past, but without any success. Very recently, Alecian et al.\ (\cite{alecian09}) discovered low amplitude (less than 1.6\,mmag) periodic variations (4.3 and 2.53~days respectively, with harmonics) in two candidate HgMn stars by means of the high quality light curves provided by the CoRoT satellite. These variations are compatible with theoretically predicted pulsation periods. However, as stated by the authors, only spectroscopic datasets could help to conclusively establish or withdraw this pulsation interpretation. Currently, there is thus no observational proof of pulsation in HgMn stars. 

The aspect of inhomogeneous distribution of some chemical elements over the surface of HgMn stars was for the first time discussed by Hubrig \& Mathys (\cite{hubrig_mathys95}). From the survey of HgMn stars in close spectroscopic binaries (SBs) it was suggested that some chemical elements might be inhomogeneously distributed on the surface, with in particular a preferential concentration of Hg along the equator. In close double-lined systems (SB2s), where the orbital plane has a small inclination to the line of sight, a rather large overabundance of Hg was found. By contrast, in stars with orbits almost perpendicular to the line of sight, mercury is not observed at all. The first indication of variability of the \ion{Hg}{II} 3984\,\AA{} and \ion{Y}{II} 3983\,\AA{} lines was reported for the HgMn SB2 system \object{AR\,Aur} by Takeda et al.\ (\cite{Takeda79}). Later, Wahlgren et al.\ (\cite{wahlgren01}) and Adelman et al.\ (\cite{adelman02}) showed that the \ion{Hg}{II} 3984\,\AA{} line of the primary component of \object{$\alpha$~And} varies with a 2.8-d period. The spectral line variations were attributed to the surface inhomogeneous mercury distribution along the stellar equator, together with the stellar rotation period. 

Recently, Kochukhov et al.\ (\cite{kochukhov05}) found clear signatures of surface mercury spots in two rapidly rotating HgMn stars by analysing the \ion{Hg}{II} 3984\,\AA{} line profiles. Variability of spectral lines associated to larger number of chemical elements were discovered for the first time by Hubrig et al.\ (\cite{hubrig06a}) for the primary component of the eclipsing binary \object{AR Aur}. The strongest variations were found for the chemical elements Pt, Hg, Sr, Y, Zr, He, and Nd. The first Doppler maps for the elements Mn, Sr, Y, and Hg were recently presented by Savanov et al.\ (\cite{sav09}). The study of Hubrig et al.\ (\cite{hubrig08}) suggests that spectral variability of various chemical elements is indeed observed in most HgMn stars.   

We present the first observational study based on a substantial number of spectra, more than one hundred, obtained with the CORALIE \'echelle spectrograph attached to the 1.2m Leonard Euler telescope in La Silla in Chile. The selected targets, the single-lined (SB1) spectroscopic binaries \object{HD\,11753} ($\phi$\,Phe, V = 5.1 mag, B\,9p) and \object{HD\,53244} ($\gamma$\,CMa, V = 4.1 mag, B8\,II), and the star \object{HD\,221507} ($\beta$\,Scl, V = 4.4 mag, B9.5\,IVmnpe) were chosen as the brightest known southern HgMn stars visible during the periods of observation. The goal of the presented spectroscopic study was twofold: to search for stellar pulsations and/or surface inhomogeneous distribution of various chemical elements. Our observations and data reduction are presented in the appendix A. 

\begin{table}
\caption[]{Atmospheric parameters (effective temperature, logarithm of the gravity, microturbulent velocity, and projected rotational velocity) taken from Dolk et al.\ (\cite{dolk2003}) for \object{HD\,11753} and \object{HD\,221507}, and from Woolf \& Lambert (\cite{woolf_lambert99}) for \object{HD\,53244}.}
\begin{center}
\begin{tabular}{ccccc}
\hline \hline\\[-7pt]
          & T$_{\rm eff}$ (K) & $\log g$ & $\xi$ (km s$^{-1}$) & $v \sin i$ (km s$^{-1}$) \\[5pt]
\hline\\[-7pt]
\object{HD\,11753}  & 10\ 612$\pm$200         & 3.79$\pm$0.10      & 0.5 $\pm$0.5               & 14$\pm$0.5\\  
\object{HD\, 53244} & 13\ 600$\pm$200         & 3.40$\pm$0.03       & 2.0 $\pm$0.5               & 35$\pm$0.5\\
\object{HD\,221507} & 12\ 476$\pm$200        & 4.13$\pm$0.10       & 0.0 $\pm$0.5               & 25$\pm$0.5\\[5pt]
\hline
\end{tabular}
\end{center}
\label{par}
\end{table} 

\begin{figure}
\centering
\includegraphics[angle=270,totalheight=0.27\textwidth]{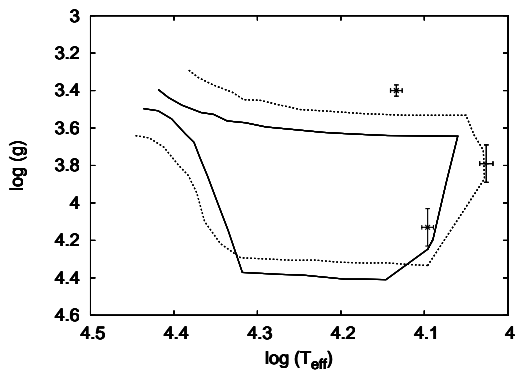}
\caption{Position of the studied stars in a \mbox{($\log T_{\rm eff}$,$\log g$)}-diagram. From left to right, the error boxes represent \object{HD\,53244}, \object{HD\,221507}, and \object{HD\,11753}. The full and dashed lines represent the boundary of the theoretical SPB instability strip for a metallicity $Z=0.02$ and $Z=0.01$, respectively (taken from Miglio et al.\ \cite{miglio07}).}
\label{HR}
\end{figure}
%

\begin{figure}
\centering
\includegraphics[angle=270,totalheight=0.35\textwidth]{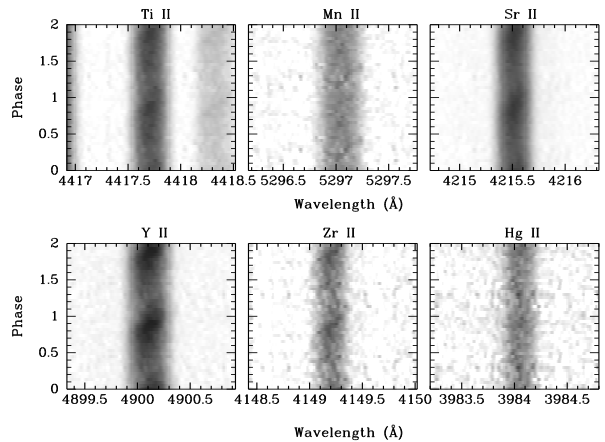}
\caption{
Spots on the surface of \object{HD\,11753}: time-series spectra phased on the stellar 
rotation period of 9.54~days. }
\label{HD11753}
\end{figure}

\begin{figure}
\centering
\includegraphics[bb= 50 30 527 720, width=4cm, height=7cm, angle=270]{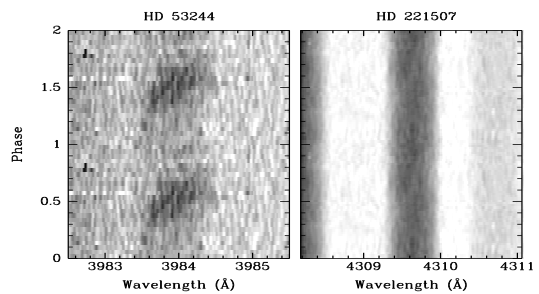}
\caption{Left: Inhomogeneous distribution of \ion{Hg}{ii} on the surface of \object{HD\,53244} apparent in the time-series spectra around \ion{Hg}{ii} $\lambda$3984 phased on the rotation period of 6.16 days. Right: Inhomogeneous distribution of \ion{Y}{ii} on the surface of \object{HD\,221507} apparent in time-series spectra around the spectral line of \ion{Y}{ii} $\lambda$4310 phased on the rotation period of 1.93 days.}
\label{HD53244}
\end{figure}
\section{Line profile variability}
Atmospheric parameters and abundances of several elements of selected HgMn stars were studied in the past by Smith \& Dworetsky (\cite{smith_dworetsky1993}), Woolf \& Lambert \ (\cite{woolf_lambert99}) and Dolk et al.\ (\cite{dolk2003}). The stellar parameters for our three studied stars are given in Table~\ref{par}. Their position in a ($\log T_{\rm eff}$,$\log g$)-diagram is presented in Fig.~\ref{HR}. The star \object{HD\,221507} is located in the SPB instability strip, whereas \object{HD\,11753} and \object{HD\,53244} are located outside this strip. According to theoretical models (e.g. Miglio et al.\ \cite{miglio07}) \object{HD\,221507} is expected to exhibit SPB-like pulsations. To search for variability we studied the behaviour of the radial velocities and equivalent widths of the spectral lines of various elements. We used the Lomb-Scargle method (Scargle\ \cite{scargle82}) and the phase dispersion minimization method (Stellingwerf \cite{stell78}). Additionally, we performed a two-dimensional frequency analysis, which is available in FAMIAS (Frequency Analysis and Mode Identification for Asteroseismology) (Zima \cite{zima08}).

Among the studied stars, the spectra of \object{HD\,11753} exhibit the most prominent variability. This was most clearly detectable for the elements Ti, Sr, and Y. The Fe lines, by contrast, show no profile variations. For that reason, we determined the stellar radial velocity (RV) by measuring 16 Fe lines free of blends. The star \object{HD\,11753} is known as a single-lined binary star, but its orbital parameters are not well known. According to our observations the orbital period would be long. In fact, all our RV measurements, which extend over 10 months, range from 14.1 to 14.5 km\,s$^{-1}$. In particular for the two first datasets (76 and 28 observations, respectively) the mean RVs are 14.24 $\pm$ 0.01 km\,s$^{-1}$ (RMS (root mean square)=0.05 km\,s$^{-1}$) and 14.36 $\pm$ 0.01 km\,s$^{-1}$ (RMS=0.05 km\,s$^{-1}$). Considering the small dispersion and the absence of any trend with the rotational velocity found from Y lines, we adopted these values as the stellar RV for the observations of the corresponding datasets. For \object{HD\,53244} our RVs present an RMS of 0.2 km\,s$^{-1}$ without significant differences between different runs. Therefore we did not apply an RV correction other than the subtraction of the mean RV.

For \object{HD\,11753}, the radial velocities and equivalent widths of the Ti, Sr, and Y lines were found to vary with the period P=9.54\,d. In the spectra of \object{HD\,53244} the variations are apparent for the elements Hg and Mn with the P=6.16\,d, while in the spectra of \object{HD\,221507} variations associated to Hg, Mn, and Y are detected, indicating the period P=1.93\,d. 

For all three observed stars the behaviour of the line profiles is different for different elements. Moreover, spectral lines associated to certain elements, like silicon, are constant. These characteristics do not support the SPB-like pulsation interpretation (De Cat \ \cite{decat01}). We thus conclude that the determined periods correspond to rotation periods of studied HgMn stars and the detected line profile variability is caused by inhomogeneous distribution of chemical elements on their stellar surface. Apart from two other HgMn stars, \object{$\alpha$~And} and \object{AR\,Aur}, no other rotation period determinations for HgMn stars were published before this study. In Figs.~\ref{HD11753} and \ref{HD53244} we display time-series spectra phased on the stellar rotation periods around selected lines of various elements in all three stars. The images were created by averaging all the spectra within the phase interval $\phi \pm 0.04$ for each phase $\phi$.

\begin{figure}
  \centering
  \includegraphics[width=0.45\textwidth]{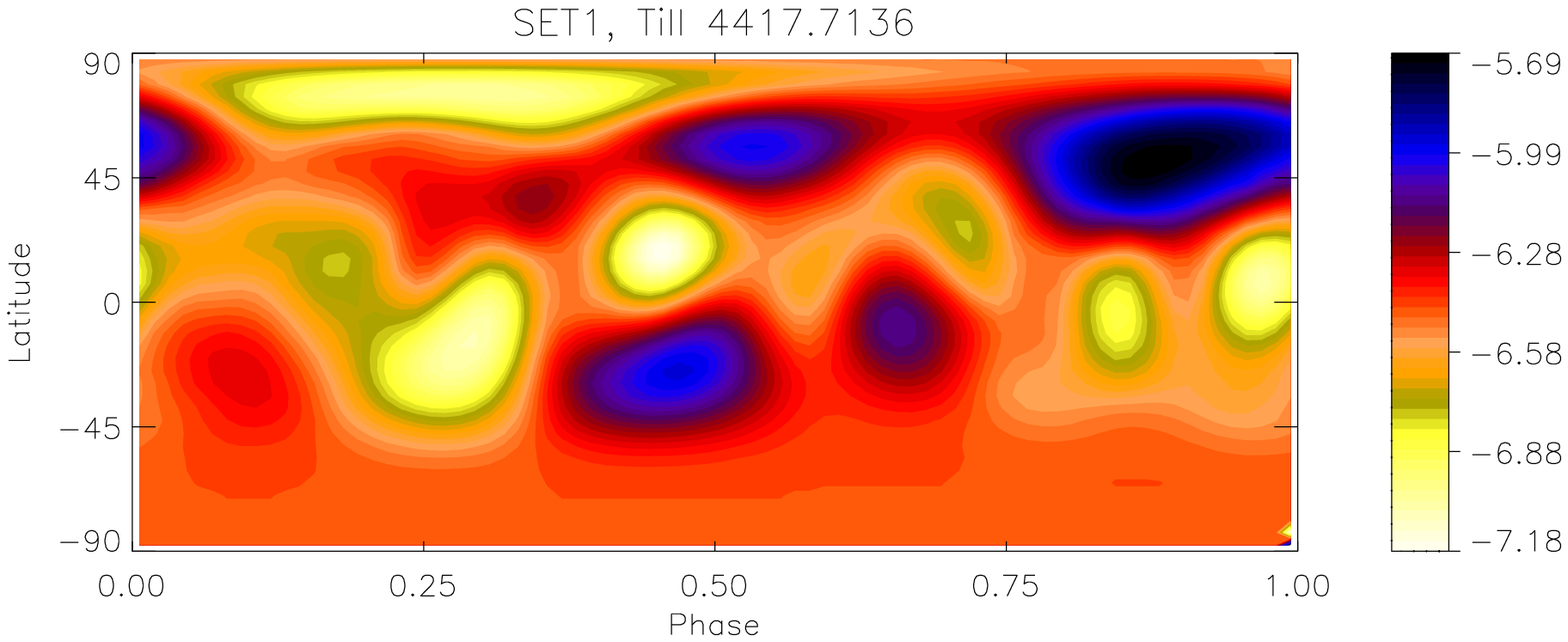}
  \includegraphics[width=0.45\textwidth]{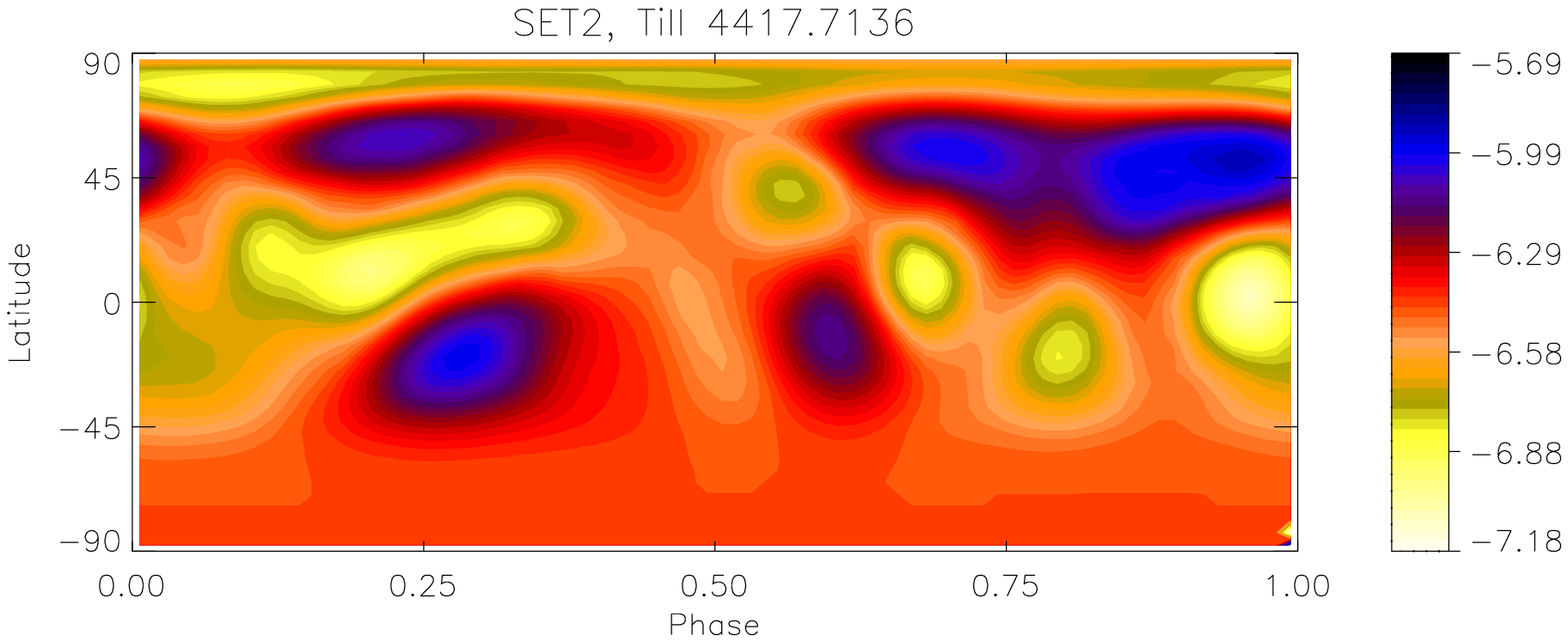}
  \caption{Ti abundance map of \object{HD\,11753} obtained from \ion{Ti}{ii} line 4417.7136\,\AA{} for set1 and set2. The colour indicates the abundance with respect to the total number density of atoms and ions. }
  \label{TiIImap}
\end{figure}

\begin{figure}
  \centering
  \includegraphics[width=0.45\textwidth]{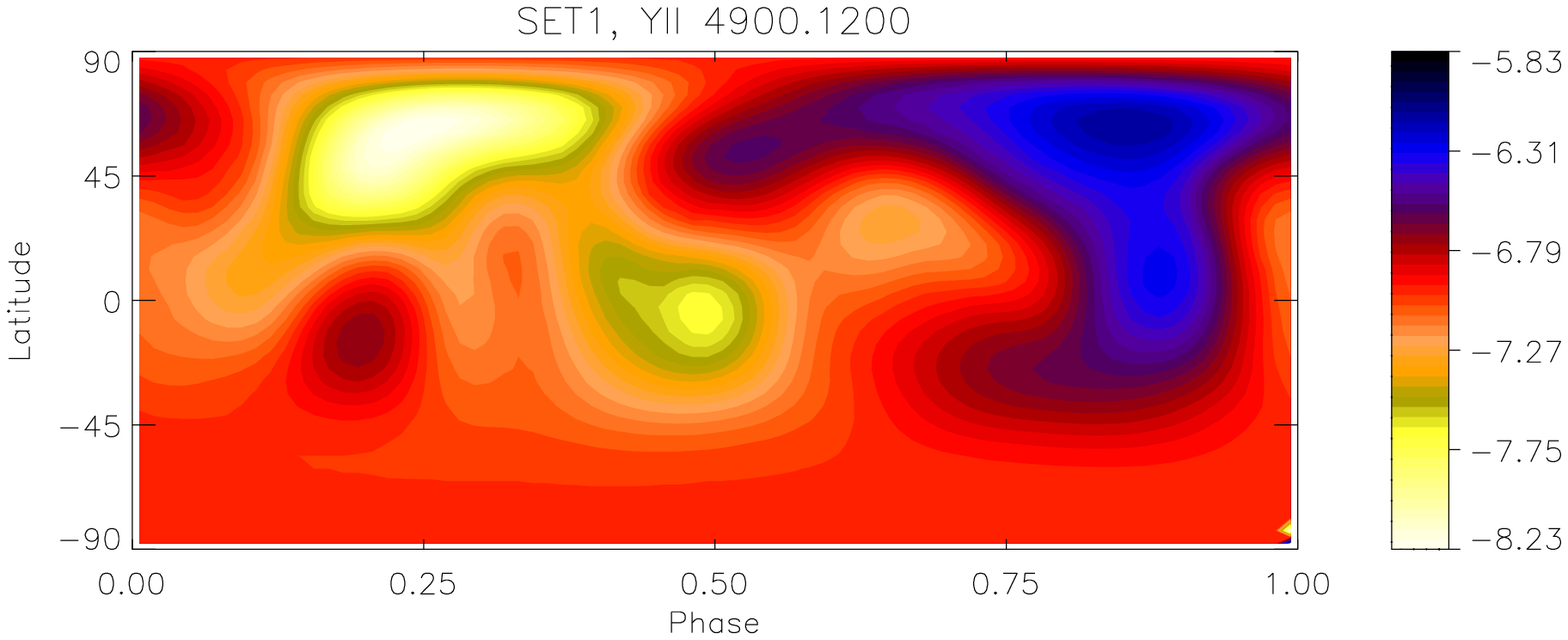}
  \includegraphics[width=0.45\textwidth]{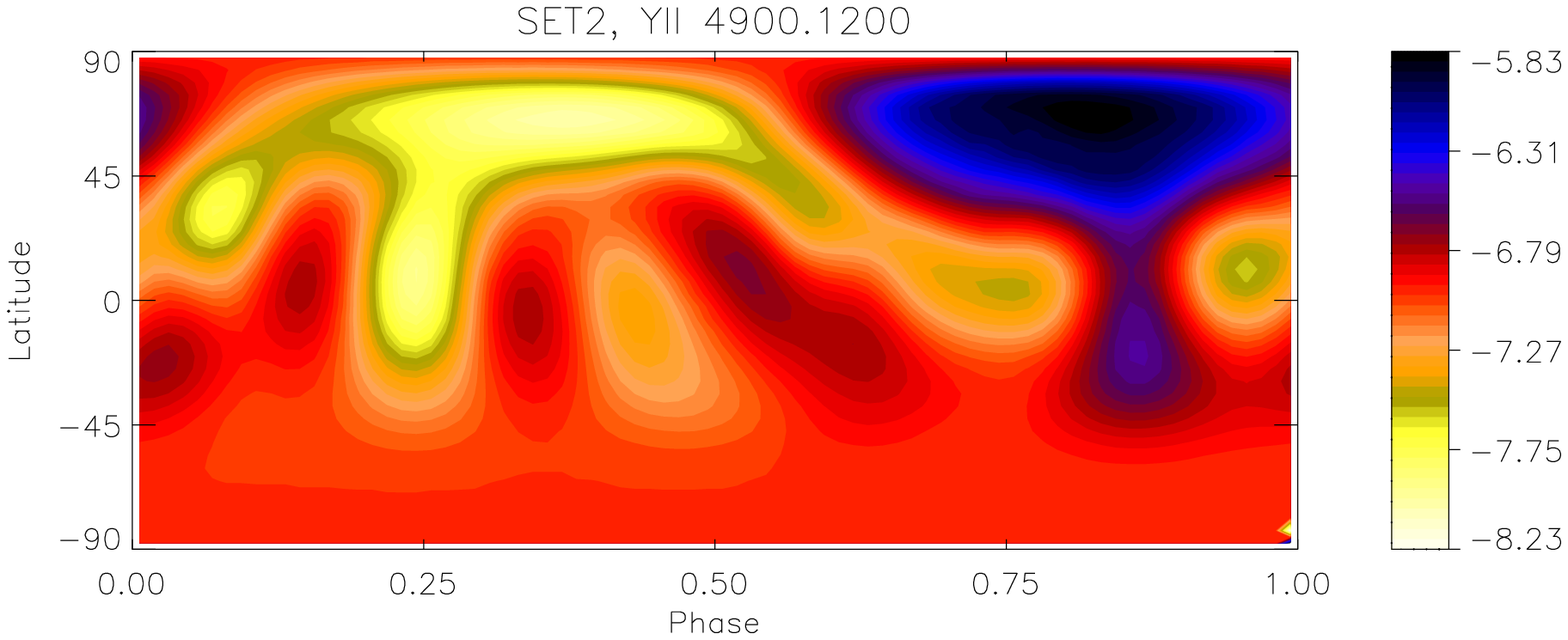}
  \caption{As in Fig.~\ref{TiIImap}, but now using the \ion{Y}{ii} line 4900.1200\,\AA{}.}
  \label{YIImap}
\end{figure}

\begin{figure}
  \centering
  \includegraphics[width=0.45\textwidth]{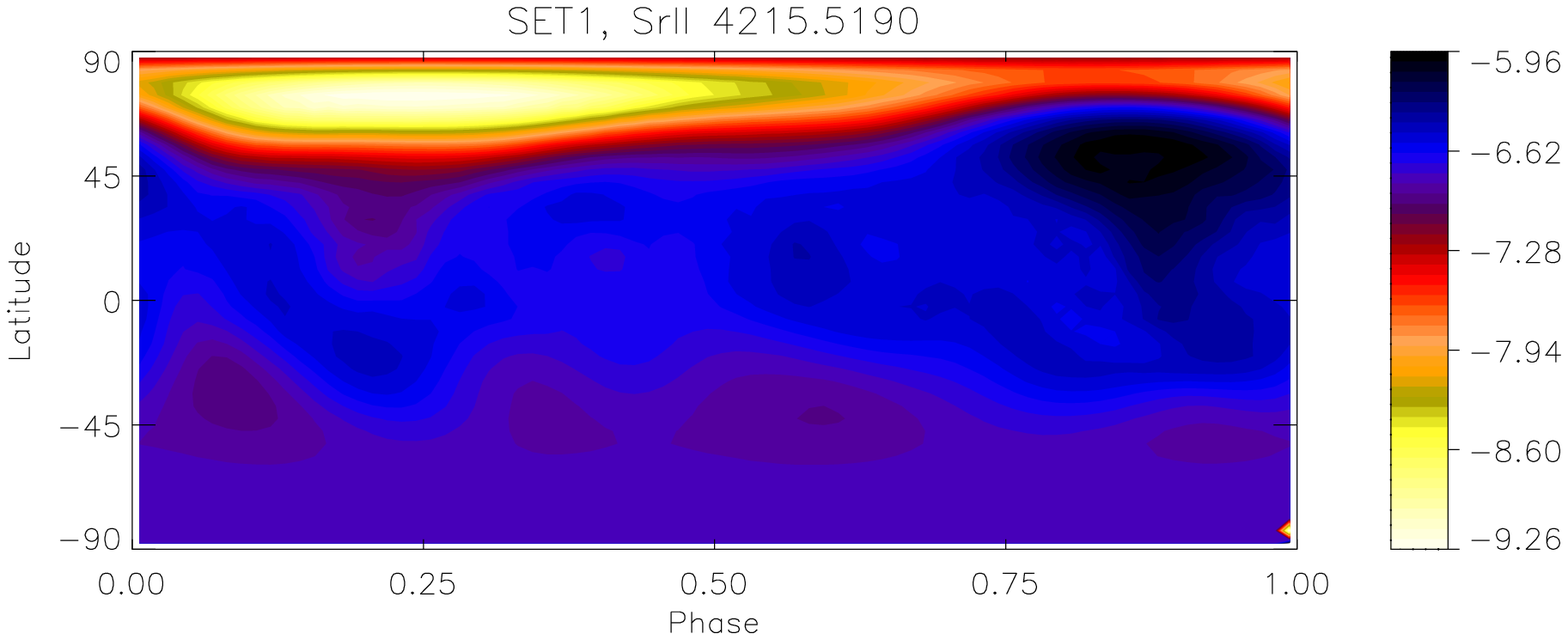}
  \includegraphics[width=0.45\textwidth]{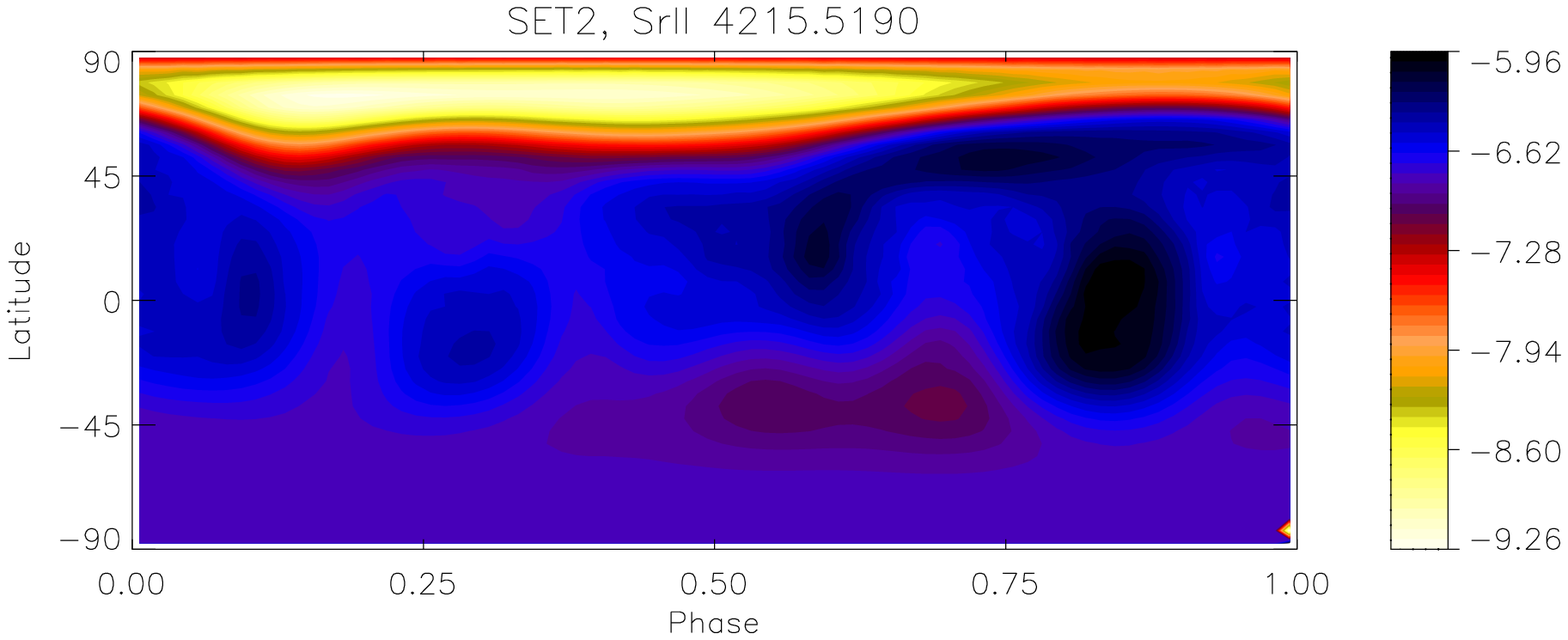}
  \caption{As in Fig.~\ref{TiIImap}, but now using the \ion{Sr}{ii} line 4215.5190\,\AA{}.}
  \label{SrIImap}
\end{figure}

\section{Surface chemical inhomogeneities by Doppler imaging reconstruction} 
For the star \object{HD\,11753}, which exhibits the most distinct spectral line profile variations for several elements, we used the Doppler imaging technique to reconstruct the surface distributions of Ti, Sr, and Y. The elements Ti and Y have numerous transitions in the observed optical spectral region allowing us to select unblended spectral lines which show strong variability.

The two sets of observations of \object{HD\,11753} obtained in 2000 Sep 28 -- Oct 11 (set1) and in 2000 Dec 02 -- Dec 15 (set2) consist of 76 and 28 observations, respectively, evenly spread over the stellar rotation cycle. Using two separate surface abundance maps based on observations of two data sets that are on average 65 days apart allowed us to obtain important information on the temporal evolution of elemental surface inhomogeneities. The Doppler imaging technique takes advantage of the partial resolution of the stellar surface provided by the rotational Doppler effect, inverting a line profile time series into a 2-D map of the stellar surface. The inversion of time series of spectroscopic observations is based on regularised image-reconstruction procedures implemented in the Doppler imaging code INVERS7PD written by N. Piskunov (see, e.g., Piskunov et al. \cite{piskunov}) and modified by T.\ Hackman (\cite{hackman}).

The observations were compared to a grid of local line profiles calculated with the SPECTRUM spectral synthesis code (Gray \& Corbally \cite{SPECTRUM}) and Kurucz model atmospheres (Kurucz \cite{kurucz93}). The local line profiles were calculated for 10 limb angles, and the stellar parameters were fixed to the values given in Table~\ref{par}. For inclination, v$\sin i$, and microturbulence inversion using several different values was carried out to determine the value that best fitted the observations. As best values for Ti and Y surface distribution reconstruction we obtained $v$\,sin\,$i$=13.5 km s$^{-1}$, $i$=53$^{\circ}$, and $\xi$=0.5 km s$^{-1}$. On the other hand, the best fit for the reconstruction of the Sr distribution was achieved for $v$\,sin\,$i$=12.3 km s$^{-1}$. The discrepancies in the $v$\,sin\,$i$ values are very likely related to the Sr vertical abundance stratification, which is frequently observed in chemically peculiar stars (e.g. Kochukhov et al.\ \cite{kochukhov06}). For each studied element a grid of abundances spanning from -4.0 to -9.5 was calculated, using a step of 0.5 in abundance. These abundances are in the scale used by SPECTRUM and are thus expressed with respect to the total number density of atoms and ions, and not with respect to hydrogen, with the log of the abundance of hydrogen set equal to 12.0.

The \ion{Ti}{ii} line chosen for the inversions is 4417.7136\,\AA{}, which has an excitation potential of 1.237\,eV. Chemical maps were separately recovered for both datasets (set1 and set2). As can be seen in Fig.~\ref{TiIImap}, the Ti abundance does not exhibit a distinct ring structure around the stellar equator as was found for the Hg distribution in the HgMn star \object{$\alpha$And} (Adelman et al.\ (\cite{adelman02}). The maps in general show a surface abundance that is higher than the solar abundance of Ti, -7.02, and have an average abundance of -6.47. Two main structures are well noticeable on the surface: a high abundance spot at high latitudes ($40^{\circ}-75^{\circ}$) spanning the phases 0.75--1.00, with an extension towards and beyond the phase 0.5 forming a half ring, and a lower abundance spot in the polar regions at the phases 0.1--0.5. Also, some discrete spots of lower and higher abundance appear in the equatorial region. The Ti lower abundance spot at phases 0.1--0.5 is less prominent in set2 than in set1. Also the spot configuration of the high abundance high latitude half ring of Ti at the latitudes $40^{\circ}-75^{\circ}$ changes between the two sets. 

The maps obtained from the \ion{Y}{ii} line 4900.1200\,\AA{} with an excitation potential of 1.033\,eV is presented in Fig.~\ref{YIImap}. The maps show a high abundance region at phases 0.5--1.0 extending from the latitude $45^{\circ}$ to the pole, with an extension to the equator around phase 0.8. The Y abundance distribution shows a high latitude lower abundance spot around phases 0.2--0.4, similarly to the Ti abundance maps. Some lower and higher abundance spots are also seen at the equatorial region. The average abundance of the Y maps is $-$7.01, which is significantly higher than the solar abundance of $-$9.80. We note that all the features revealed in the maps show abundances that are higher than the solar abundance of Y. Similar to the Ti maps, we observed in the Y maps that the lower abundance high latitude feature at phases 0.2--0.4 becomes more extended and less prominent in set2. Additionally, the abundance of the high abundance spot at phases 0.6--1.0 increases.

The Sr maps based on the modelling of the \ion{Sr}{ii} 4215.5190\,\AA{} line reveal similarities in the distribution of overabundant and underabundant features, though the similarity is more pronounced in the Sr and Y maps. In Fig.~\ref{SrIImap} the equatorial and polar belt-like Sr structures resemble rather well those of Y. The time evolution of the chemical spots is also seen in the Sr maps, with the lower abundance high latitude feature at phases 0.1--0.5 becoming more extended in set2, and the high abundance feature of the phases 0.8--1.0 moving towards the equator and phases 0.5. The average abundance of the Sr maps is $-$6.89, which exceeds the solar abundance of $-$9.07. 

One has to keep in mind though that set2 has a gap at the phases 0.187-0.391. This phase gap can affect the achieved resolution on the surface at these phases, and can especially hamper the exact determination of the latitudes. Still, the changes between the two sets are seen from the maps at other phases, and also from the line profiles themselves as shown in Appendix~\ref{variability}.

\section{Discussion}
All Ti, Sr, and Y abundance maps reveal a structure reminiscent of broken rings of low and high abundance. This elemental distribution is to some extent similar to the maps previously reconstructed for another HgMn star, \object{AR\,Aur} (Savanov et al.\ \cite{sav09}), where the elements Mn, Y, Sr, and Hg show abundance concentration in equatorial and polar features. Typically, inhomogeneous chemical abundance distributions are observed only on the surface of magnetic chemically peculiar stars with large-scale organised magnetic fields. In these stars, the abundance distribution of certain elements is non-uniform and non-symmetric with respect to the rotation axis. A magnetic field of the order of a few hundred Gauss was detected in hydrogen lines of four HgMn stars by Hubrig et al.\ (\cite{hubrig06b}) using low-resolution ($R=2000$) circular polarisation spectra obtained with FORS\,1 at the VLT. This small sample of HgMn stars also included the spectrum variable HgMn star \object{$\alpha$~And} with a magnetic field of the order of a few hundred Gauss. On the other hand, high-resolution spectropolarimetric spectra of some HgMn stars, including \object{$\alpha$~And}, were used in studies of Shorlin et al. (\cite{shorlin02}) and Wade et al. (\cite{wade06}), where no detection was achieved using all metal lines together in the least-squares deconvolution multi-line profile. Although strong large-scale magnetic fields have not generally been found in HgMn stars, it has never been ruled out that these stars might have tangled magnetic fields of the order of a few thousand Gauss with no net longitudinal component (e.g., Mathys \& Hubrig \cite{mat95}; Hubrig et al.\ \cite{hubrig99}; Hubrig \& Castelli \cite{hubrig01}). It is of interest that magnetohydrodynamical simulations by Arlt et al. \ (\cite{arlt03}), which combine a poloidal magnetic field and differential rotation can produce a magnetic field topology that is similar to the broken elemental ring structures seen in \object{HD\,11753} and \object{AR\,Aur}. These simulations and their implication have been recently discussed by Hubrig et al.\ (\cite{hubrig08}). 

The abundance maps of \object{HD\,11753} presented in this work exhibit clear differences between the surface abundance distribution of Ti, Sr, and Y. We also detected clear differences in the spot configurations obtained from the same lines but for different data sets, which indicates a rather fast dynamical evolution of the abundance distribution with time. Kochukhov et al.\ (\cite{kochukhov07}) discovered mercury clouds in the atmosphere of a HgMn star \object{$\alpha$~And} that showed secular changes with a time period of 2--4 years. In our analysis, using two datasets separated by $\sim$65 days, we reveal that the changes in the chemical spot configuration of \object{HD\,11753} appear much faster and can already be detected at a time scale of months. 

The results reported in this paper open up new perspectives for our knowledge and understanding of HgMn stars. Different dynamical processes take place in stellar radiation zones. An interaction between the differential rotation, the magnetic field, and the meridional circulation could possibly play a role in the generation of dynamical evolution of chemical spots. From the comparison of maps we find that it is possible that the Y and Sr distributions show indications of an increasing rotation rate towards the rotation pole, so-called differential rotation of anti-solar type. On the other hand, further analyses of the elemental surface distribution in a larger sample of HgMn stars should be carried out before the implication of these new results can be discussed in more detail.

\begin{acknowledgements}
We thank our colleagues from the Institute of Astronomy of Leuven University, who contributed to the gathering of these data.      
\end{acknowledgements}

\appendix
\section{Observations and data reduction}
\begin{figure}
  \centering
  \includegraphics[width=8cm]{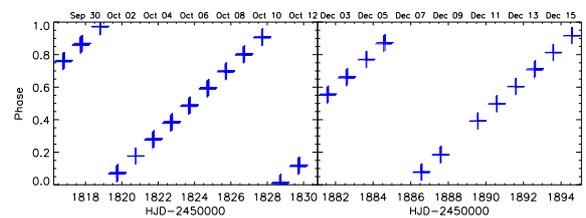}
  \caption{Phases of the \object{HD\,11753} observations for both sets (set1 on the left).}
  \label{phases}
\end{figure}

Our spectroscopic data were obtained with the CORALIE \'echelle spectrograph attached to the 1.2m Leonard Euler telescope in La Silla in Chile. In total, we gathered 113, 103, and 105 useful spectra for \object{HD\,11753}, \object{HD\,53244}, and \object{HD\,221507}, respectively. The observing logbook is given in Table~\ref{log} and the obtained phases for \object{HD\,11753} in Fig.~\ref{phases}. The wavelength domain of the CORALIE fibre-fed spectrograph ranges from 3875 to 6820\,\AA{} recorded on 68 orders. The CCD camera is a 2k x 2k CCD with pixels of 15 $\mu$m. CORALIE reaches a spectral resolution of 50\,000 with a 3 pixel sampling.

To search for variability due to pulsation in line profiles we used rather short integration times to obtain a S/N ratio of about 100 near the wavelength 4130\,\AA{}, which is the position of a prominently present \ion{Si}{II} doublet. This spectral region is well-suited for studying line profile variability of SPB stars (e.g., De Ridder et al.\ \cite{deridder02}).

We used the online reduction package available for the CORALIE spectrograph based on the method of Baranne et al.\ (\cite{baranne96}). The process involves the usual steps of de-biasing, flat-fielding, background subtraction and wavelength calibration by means of measurements of a ThAr calibration lamp. We did a more precise correction for the pixel-to-pixel sensitivity variations by using all available flat fields obtained during the night instead of using only one flat field, as is done by the on-line reduction procedure. After a correction to the heliocentric frame, the spectra were normalised using a cubic spline fit.

\begin{table}
\caption[]{Logbook of our spectroscopic observations of the three HgMn stars. N denotes the number of 
observations and the HJD is given 
with respect to HJD$_0$ = 2400000.}
\begin{center}
\begin{tabular}{cccc}
\hline \hline\\[-7pt]
& N & HJD begin & HJD end  \\[5pt]
\hline\\[-7pt]
HD\,11753 & 76 & 51816 & 51829 \\
& 28 & 51881 & 51894 \\
& 6  & 51942 & 51948 \\
& 3  & 52111 & 52123  \\[5pt]
\hline\\[-7pt]
HD\,53244 & 56 & 51816 & 51829 \\
& 18 & 51881 & 51893 \\
& 29 & 51940 & 51953 \\[5pt]
\hline\\[-7pt]
HD\,221507 & 68 & 51816 & 51829 \\
& 31 & 51881 & 51894 \\
& 6 & 52110& 52122 \\[5pt]
\hline
\end{tabular}
\end{center}
\label{log}
\end{table}

\section{Details on Doppler imaging}

\begin{figure*}
  \centering
  \includegraphics[width=12cm]{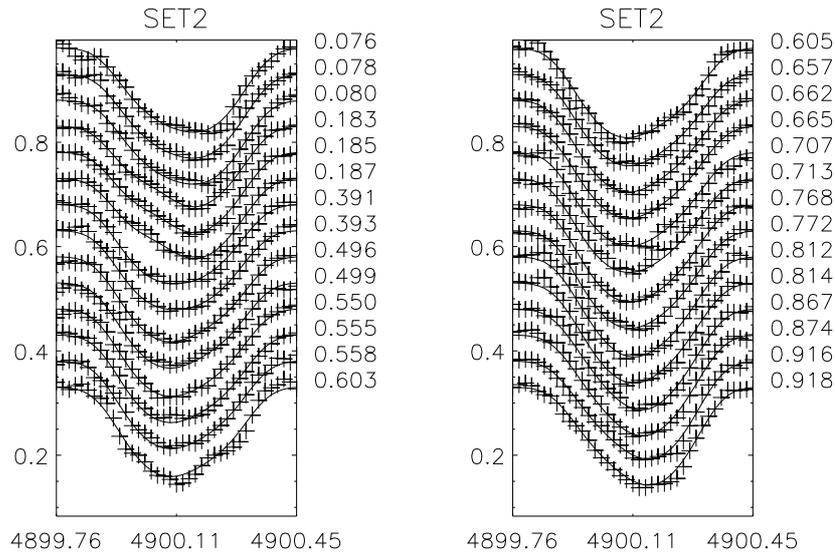}
  \caption{Set2 \ion{Y}{ii}\,4900.1200~{\AA} spectra of \object{HD\,11753} used in the Doppler imaging. Calculated and observed spectral lines are shown by lines and crosses, respectively. The numbers on the right side of the plots give the phase for each spectrum.}
  \label{linefits}
\end{figure*}

The average S/N ratio of the \object{HD\,11753} observations for set1 and set2, measured at 4400~{\AA} are 132 and 143, respectively. This means that in Doppler imaging a smaller deviation than this should not be achieved between the model and the observations. In the Tikhonov regularisation used by INVERS7PD the regularisation has to be chosen in a way that the noise in the observations is not mapped. For \object{HD\,11753} the deviation between the model and observations for the set1 is between 0.750\% (\ion{Y}{ii}\,4900.120{\AA}) and 0.893\% (\ion{Sr}{ii}4215.5190~{\AA}), and for the set2 between 0.628\% (\ion{Y}{ii}\,4900.1200{\AA}) and 0.803\% (\ion{Sr}{ii}\,4215.5190~{\AA}). These values are comparable to the average noise level in the observations. One has to also keep in mind that the noise level in the blue part is higher than in the red part of the spectrum. Furthermore, the comparison of the model and observations shows that the line profiles are well fitted without the model following the noise. An example of the line profile fits is shown in Fig.~\ref{linefits} for the \ion{Y}{ii}\,4900.1200{\AA} line of set2.

\section{Line variability}
\label{variability}

\begin{figure*}
  \centering
  \includegraphics[width=16cm]{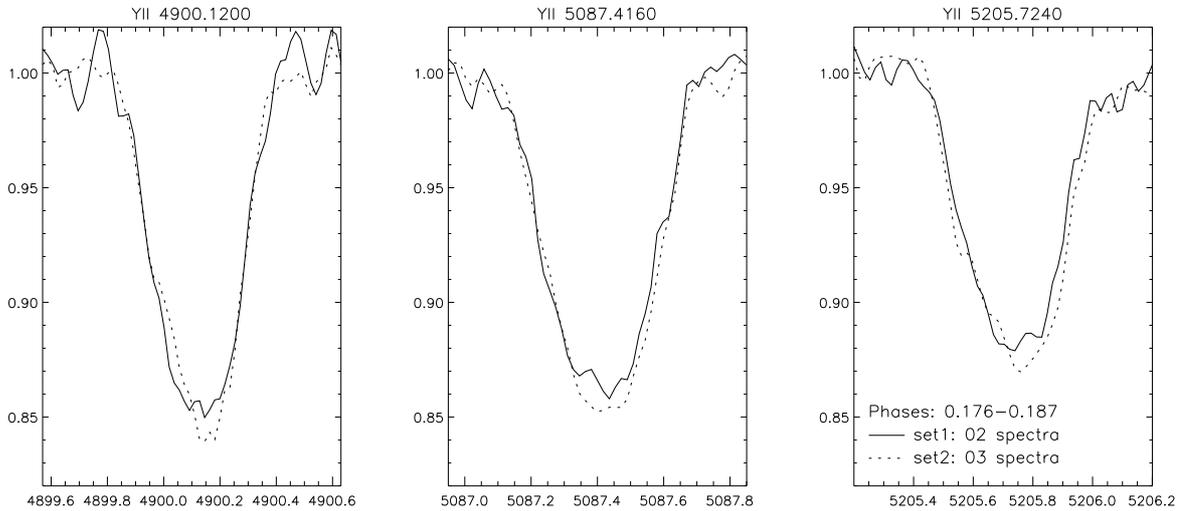}
  \caption{Mean \ion{Y}{II} line profiles from close-by phases for set1 (phases 0.176--0.177) and set2 (phases 0.183--0.187). The two mean spectra are plotted for three YII lines, which have similar excitation energies: 4900.1200 {\AA}, 5087.4160 {\AA} and YII 5205.7240 {\AA}. In the plots set1 is given by the solid line and set2 by the dashed line.}
  \label{profiles018}
\end{figure*}

The changes between the two sets are also seen when looking at the line profiles at the phases where the largest differences in the maps occur. Figure~\ref{profiles018} shows averaged spectra of very close-by phases from set1 (mean of two spectra, phases: 0.176--0.177) and set2 (mean of three spectra, phases: 0.183--0.187). The mean profiles for the two sets are shown for three \ion{Y}{II} lines of similar strength and excitation energy (4900.1200 {\AA}, 5087.4160 {\AA} and 5205.7240 {\AA}). As can be seen all three lines show changes between the two sets that are similar and in-line with the changes seen in the Doppler images, i.e., the \ion{Y}{II} abundance is higher for set2 than set1 at these phases. 


\begin{thebibliography}{}


\bibitem[1972]{abt72}Abt et al., 1972, ApJ 175, 779

\bibitem[2002]{adelman02}Adelman, S.J., Gulliver, A.F., Kochukhov, O.P., Ryabchikova, T.A. 2002, ApJ, 575, 449

\bibitem[2009]{alecian09} Alecian, G., Gebran, M., Auvergne, M., Richard, O., Samadi, R., Weiss, W.W., Baglin, A., 2009, A\&A, 506, 69

\bibitem[2003]{arlt03}
Arlt, R., Hollerbach, R.,\& R{\"u}diger, G. 2003, A\&A 401, 1087

\bibitem[1996]{baranne96}Baranne, A., Queloz, D., Mayor, M., et al., 1996, A\&A, 119, 373

\bibitem[2001]{decat01}De Cat, P. 2001, PhD thesis, KULeuven, Belgium

\bibitem[2003]{decat03}De Cat, P. 2003, Ap\&SS, 284, 37

\bibitem[2007]{decat07}De Cat, P., Briquet, M., Aerts, C., et al., 2007, A\&A 463, 243

\bibitem[2002]{deridder02}De Ridder, J., Dupret, M.-A., Neuforge, C., Aerts, C. 2002, A\&A, 385, 572

\bibitem[2003]{dolk2003} Dolk, L., Wahlgren, G.M., \& Hubrig, S., 2003, A\&A 402, 299

\bibitem[1994]{SPECTRUM} Gray, R.O.,\& Corbally, C.J. 1994, AJ, 107, 742

\bibitem[2001]{hackman} Hackman, T., Jetsu, L.,\& Tuominen, I. 2001, A\&A, 374, 171

\bibitem[1995]{hubrig_mathys95}Hubrig, S., Mathys, G. 1995, Com. Ap, 18, 167

\bibitem[1999]{hubrig99}Hubrig, S., Castelli, F., \& Wahlgren, G.M., 1999, A\&A 346, 139

\bibitem[2001]{hubrig01}Hubrig, S. \& Castelli, F., 2001, A\&A 375, 963

\bibitem[2006a]{hubrig06a}Hubrig, S., Gonz\'alez, J.F., Savanov, I., Sch\"oller, M., Ageorges, N., Cowley, C.R., Wolff, B. 2006a, MNRAS, 371,1953

\bibitem[2006b]{hubrig06b}Hubrig, S., North, P., Sch\"oller, M., \& Mathys, G., 2006b, AN 327, 289

\bibitem[2008]{hubrig08}Hubrig, S., Gonz\'alez, J.F., Arlt, R. 2008, CoSka, 38, 415

\bibitem[2005]{kochukhov05}Kochukhov, O., Piskunov, N., Sachkov, M., Kudryavtsev, D., 2005, A\&A, 439, 1093

\bibitem[2006]{kochukhov06}Kochukhov, O., Tsymbal, V., Ryabchikova, T., Makaganyk, V., Bagnulo, S., 2006, A\&A, 460, 831

\bibitem[2007]{kochukhov07}Kochukhov, O., Adelman, S.J., Gulliver, A.F.,\& Piskunov, N. 2007, Nature Physics, 3, 526

\bibitem[1993]{kurucz93} Kurucz, R.L., 1993, Kurucz CD No. 13

\bibitem[1995]{mat95}Mathys, G. \& HUbrig, S., 1995, A\&A, 293, 810

\bibitem[2007]{miglio07} Miglio, A., Montalb\'an, J., Dupret, M-A., 2007, MNRAS, 375, 21

\bibitem[1990]{piskunov}Piskunov, N.E., Tuominen, I.,\& Vilhu, O. A\&A, 230, 363

\bibitem[2009]{sav09} Savanov, I.S., Hubrig, S., Gonz\'alez, J. F., \& Sch\"oller, M., 2009, IAU Symp. v. 259, p. 401

\bibitem[1982]{scargle82} Scargle, J.D. 1982, ApJ 263, 835

\bibitem[1993]{smith_dworetsky1993}Smith, K.C., Dworetsky, M.M. 1993, A\&A 274, 335

\bibitem[2002]{shorlin02} Shorlin, S. L. S., Wade, G. A., Donati, J.-F. et al., 2002, A\&A 392, 637

\bibitem[1978]{stell78} Stellingwerf, R.F., 1978, ApJ 224, 953

\bibitem[1979]{Takeda79} Takeda Y., Takada M., Kitamura M., 1979, PASJ, 31, 821

\bibitem[2005]{turcotte_richard05} Turcotte, S., Richard, O. 2005, EAS Publications Series, Vol.17, pp.357--360

\bibitem[2006]{wade06} Wade, G. A., Auriere, M., Bagnulo, S. et. al., 2006, A\&A 451, 293

\bibitem[2001]{wahlgren01}Wahlgren, G.M., Ilyin, I., Kochukhov, O. 2001, AAS, 33, 1506

\bibitem[1999]{woolf_lambert99}Woolf, V.M., Lambert, D.L. 1999, ApJ, 521, 414

\bibitem[2008]{zima08} Zima, W. 2008, CoAst, 157, 387
\end{thebibliography}
\end{document}